\documentclass[sigconf,authorversion,nonacm]{acmart}
\AtBeginDocument{%
  }

\setcopyright{acmlicensed}
\copyrightyear{2024}
\acmYear{2024}
\acmDOI{XXXXXXX.XXXXXXX}

\acmISBN{978-1-4503-XXXX-X/18/06}

\begin{document}

\title[Integrating LLMs in Education: Challenges and Prospects]{The Future of Learning: Large Language Models through the Lens of Students}

\author{He Zhang}
\email{hpz5211@psu.edu}
\affiliation{%
  \institution{College of Information Sciences and Technology, Pennsylvania State University}
  \city{University Park}
  \state{Pennsylvania}
  \country{USA}
  \postcode{16802}
}
\author{Jingyi Xie}
\email{jzx5099@psu.edu}
\affiliation{%
  \institution{College of Information Sciences and Technology, Pennsylvania State University}
  \city{University Park}
  \state{Pennsylvania}
  \country{USA}
  \postcode{16802}
}
\author{Chuhao Wu}
\email{cjw6297@psu.edu}
\affiliation{%
  \institution{College of Information Sciences and Technology, Pennsylvania State University}
  \city{University Park}
  \state{Pennsylvania}
  \country{USA}
  \postcode{16802}
}
\author{Jie Cai}
\email{jpc6982@psu.edu}
\affiliation{%
  \institution{College of Information Sciences and Technology, Pennsylvania State University}
  \city{University Park}
  \state{Pennsylvania}
  \country{USA}
  \postcode{16802}
}
\author{ChanMin Kim}
\email{cmk604@psu.edu}
\affiliation{%
  \institution{College of Education, Pennsylvania State University}
  \city{University Park}
  \state{Pennsylvania}
  \country{USA}
  \postcode{16801}
}
\author{John M. Carroll}
\email{jmc56@psu.edu}
\affiliation{%
  \institution{College of Information Sciences and Technology, Pennsylvania State University}
  \city{University Park}
  \state{Pennsylvania}
  \country{USA}
  \postcode{16802}
}

\renewcommand{\shortauthors}{Zhang et al.}

\begin{abstract}
As Large-Scale Language Models (LLMs) continue to evolve, they demonstrate significant enhancements in performance and an expansion of functionalities, impacting various domains, including education. In this study, we conducted interviews with 14 students to explore their everyday interactions with ChatGPT. Our preliminary findings reveal that students grapple with the dilemma of utilizing ChatGPT's efficiency for learning and information seeking, while simultaneously experiencing a crisis of trust and ethical concerns regarding the outcomes and broader impacts of ChatGPT. The students perceive ChatGPT as being more ``human-like'' compared to traditional AI. This dilemma, characterized by mixed emotions, inconsistent behaviors, and an overall positive attitude towards ChatGPT, underscores its potential for beneficial applications in education and learning. However, we argue that despite its human-like qualities, the advanced capabilities of such intelligence might lead to adverse consequences. Therefore, it's imperative to approach its application cautiously and strive to mitigate potential harms in future developments.
\end{abstract}




\maketitle

\section{Introduction}
\label{sec:intro}
In the swiftly evolving landscape of artificial intelligence (AI), large-scale language models (LLMs) have emerged as a pivotal force for innovation and a driving force behind next-generation efficiency. LLMs, with their seemingly ``omnipotent'' capabilities, especially traits of general-purpose technologies~\citep{eloundou2023gpts}, are increasingly performing at levels comparable to humans in various tasks~\cite{teubner2023welcome}. Their remarkable performance enhancements, coupled with an expanding range of functionalities—including text processing, question-answer dialogues, programming, image interpretation, and video creation—have captivated both academic and industrial communities. These advancements are progressively convincing the broader public of the significance of these models~\citep{sanderson2023gpt}. Understanding and utilizing LLMs are becoming essential skills in our daily lives~\citep{dell2023navigating}, akin to the use of personal computers and smartphones.

Beyond comparisons with human capabilities, LLMs, due to their vast training data scale, complexity, more comprehensive understanding of tasks, and versatility in various scenarios are replacing some traditional AI tools. They surpass these traditional tools in a range of tasks, including writing~\citep{zhang2023sirens}, diagnosis~\citep{wang2023llms}, and retrieval~\citep{davis2023evaluating}. This advancement illustrates that LLMs are actively redefining our perceptions of future possibilities in diverse fields. As ~\citet{lee2021ai} states in their book, this powerful AI will have an impact on areas including autonomous driving, career choices, virtual companions, education, ethical concepts, and broader social issues. 

Among these, education is at the forefront of discussion, particularly since the COVID-19 pandemic. Education, long considered a crucial aspect of societal welfare~\citep{hanushek2010education}, has undergone several significant transformations and challenges, from campus social distancing~\citep{10.1145/3411763.3451526} and virtual classrooms~\citep{10.1145/3411764.3445240} to independent study~\citep{10.1145/3411764.3445450} and collaborative learning~\citep{10.1145/3491102.3517641}. In the context of the rapid development of LLMs, education, due to its importance, has naturally become one of the first fields to be impacted by such technologies, with related issues attracting increasing attention from researchers. Prior to this, the impact of AI on education had already been a topic of widespread discussion in the HCI community~\citep{KASNECI2023102274}. 

In this study, we focused on ChatGPT, a prime example of LLMs. We conducted semi-structured interviews with 14 participants from diverse educational and professional backgrounds to gain valuable insights into their experiences with LLMs' applications. By collating and analyzing their perspectives, we further elucidate the practical challenges and opportunities encountered in the utilization of LLMs. We discuss the potential opportunities to leverage LLMs in the education context. 

Specifically, this study aims to address the following research questions:
 \begin{description}
    \item[RQ1.] What are the impacts and scenarios of using ChatGPT on Intentional Learning and Incidental Learning?
    \item[RQ2.] What are the attitudes towards collaborative learning with LLMs from students' perspective?
\end{description}

By exploring these questions, we seek to understand how ChatGPT and similar technologies can be integrated into educational settings to enhance learning outcomes and foster a more interactive and efficient learning environment.

\section{Related Work - AI in Education}
LLMs are seen as valuable learning tools in educational settings now. For example, Kazemitabaar et al.~\citep{10.1145/3613904.3642773} developed CodeAid, an LLM-powered programming assistant, and found that CodeAid significantly influenced student engagement and learning, highlighting distinct usage patterns and the effectiveness of responses in aiding programming tasks. Jin et al.~\citep{10.1145/3613904.3642349} examined the use of LLMs as teachable agents in programming education. While their study showed benefits in knowledge-building and metacognitive skills, challenges with authentic interactions and knowledge transfer were noted. Han et al.~\citep{10.1145/3613904.3642438} highlighted the potential benefits of AI in providing adaptive teaching materials and personalized feedback, while also addressing significant concerns about authorship, agency, and misinformation. These insights underscore the need for careful design and regulation of educational AI platforms. Shaer et al.~\citep{10.1145/3613904.3642414} examined the use of LLMs in group ideation within educational settings. Their research showed that LLMs could enhance creativity and support collaborative innovation, especially during the idea generation process. However, the study also pointed out the limitations and biases of non-human agents in evaluating ideas.

These studies collectively contribute to understanding the impact of LLMs in education, demonstrating their potential to enhance learning and creativity while also highlighting the importance of addressing trust, dependency, and ethical concerns~\citep{Teogivensage2024}. However, we found that previous research primarily emphasizes the advantages of LLMs as tools, highlighting their effectiveness~\citep{10.1145/3613904.3642773,10.1145/3613904.3642349,zhang2023qualigpt}, or focuses on attitudes towards their use~\citep{10.1145/3613904.3642414,wureacting}, such as concerns or trust~\citep{10.1145/3613904.3642438}, and discussions on learning models remain relatively underexplored. Therefore, while our findings align with the overall trend in the literature, that LLMs are effective but can be optimized further, we specifically address research questions related to learning models, especially intentional learning and incidental learning.

 Overall, while there are some concerns, there is a generally positive attitude toward the use of LLMs in education~\citep{10.1145/3613904.3642773,HADIMOGAVI2024100027}.

\section{Methods}

\subsection{Participants Recruitment}

We recruited 14 participants (P1-P14) through social media and the authors' network of contacts, including 11 females and 3 males; 1 undergraduate student, 1 master's student, and the rest were PhD students.The age range of the participants was 18 to 35 years (median = 27, SD~$\approx$~3.08). The study was conducted under the approval of the university's Institutional Review Board (IRB). At the conclusion of the study, each participant received a \$10 gift card (or an equivalent amount) as compensation. 


\subsection{Data Collection and Analysis}
This study was conducted online through video conferencing software (e.g., Zoom). The research process was recorded after the informed consent of all participants.
The interviews were semi-structured, lasted from 30 minutes to 1 hour. Initially, we asked participants to introduce their professional experiences and background information. Subsequently, we delved into the main challenges they encountered while using ChatGPT. Finally, we discussed the potential of integrating ChatGPT and other LLMs into the educational sector with the participants. Through thorough review, analysis, and reflection on the recorded sessions and their respective codings, we unearthed insights regarding participants' experiences using ChatGPT, as well as their attitudes towards applying ChatGPT or other LLMs in the field of education. 

We conducted reflexive thematic analysis (RTA) on the collected data~\citep{clarke2015thematic}. The data analysis generally adhered to a six-step procedure: dataset familiarization, data coding, initial theme generation, theme development and review, theme refinement and definition, and report composition. After each interview and experiment, the research team briefly discussed the outcomes. All recordings were transcribed and coded by the first author and at least one other author. During this research process, the researchers met at least once a week to discuss the progress of the study, the results of the interviews and experiments, the findings and problems, and to continually refine the themes and processes.

\section{Findings}

\subsection{Intentional Learning through ChatGPT}
Intentional learning through ChatGPT embodies a focused and purposeful approach to acquiring knowledge or skills. In this context, participants engage with this tool, aiming for specific goals: to obtain targeted information more efficiently and clearly than traditional search engines, or to spark creative thinking.

\subsubsection{ChatGPT as an Alternative to Traditional Search Engines}
Participants have shifted from traditional search engines like Google to posing questions directly to ChatGPT, which has significantly improved their efficiency in information retrieval. These inquiries often pertain to established concepts, theories, or general information.

One of the most notable benefits of ChatGPT is its ability to rapidly synthesize and integrate complex information from multiple sources. This capability significantly surpasses traditional search engines in terms of speed and convenience. The efficiency of ChatGPT not only streamlines information retrieval but also impacts user reliance and search behavior.
 \begin{quote}
     \textit{``When I have questions, I don't really want to use search engines now. I feel that when I ask something, it [ChatGPT] gives me a direct answer, so I don't need to search through numerous responses [provided by search engines] to find what I want. Moreover, I think the results given by both [ChatGPT and search engines] are not too different.''} (P5)
 \end{quote}

In addition to its efficiency in information retrieval, ChatGPT excels in breaking down intricate concepts into more understandable terms, such as in the fields of finance and economics. This capability aids in immediate understanding for individuals without a background in these fields and makes specialized knowledge accessible to a broader audience. 
 \begin{quote}
     \textit{``For example, for some knowledge about finance and economics, the search might be very complex, and very hard to understand. So, at this time, I use ChatGPT. I think that it must have seen these, its database includes these contents. I can ask it to explain complex concepts in an accessible manner.''} (P9)
 \end{quote}

Although ChatGPT can provide rapid and effective responses, participants have significant hesitation in fully trusting its output, primarily due to concerns about accurcay and potential content fabrication. This is because, fundamentally, ChatGPT \textit{``does not search like a search engine itself searches''} (P9), but rather generates responses based on its training data. 
The accuracy and bias of these responses are significant points of concern, as highlighted by P12: \textit{``I guess the only concern I have is that I don't know if it's really accurate.''} Users have encountered inaccuracies and overly literal responses, leading them to rephrase questions or restart interactions for clarity. This lack of confidence in the verifiability of ChatGPT’s results, especially in the absence of source information, also extends to its effectiveness in subjective tasks and concerns about its randomness and unstructured responses.

\subsubsection{ChatGPT as a Mentor to Inspire Ideas}
Another advantage of ChatGPT in intentional learning is its role in fostering inspiration. Particularly in situations where creative ideas are needed, ChatGPT can quickly provide insights to stimulate users' thinking. For instance, P5 described integrating ChatGPT into the design process: \textit{``When we are in the design process, especially during the initial brainstorming phase, we incorporate ChatGPT into this process, asking it to see if it can provide any good inspiration.''} 
This approach is not unique to design. P7 found ChatGPT helpful to \textit{``draft an outline for interview questions''}. 
Similarly, P8 used ChatGPT to brainstorm possible topics and titles for writing tasks. These instances highlight the role of ChatGPT in aiding the generation of creative ideas and enriching the brainstorming phases. 

\begin{quote}
    \textit{``Sometimes, when I feel that my writing wasn't good, I ask ChatGPT to help revise it or give me some suggestions for possible topics. Also, when I'm not sure what title I should use for an article, I let it [ChatGPT] give me a possible title, and then I use this title.''} (P8)
\end{quote}

A major concern with the increasing reliance on tools like ChatGPT is the potential erosion of critical thinking and learning abilities. As these technologies take over tasks traditionally requiring human cognition, there is a risk of individuals becoming overly dependent on AI for problem-solving and creative thinking. P11 succinctly captures this apprehension: \textit{``I feel that everyone might gradually lose their ability to think, as people won't be willing to think anymore. They'll just rely on machines to do the thinking for them.''}

This concern is further echoed in the context of younger learners. P9 highlighted the potential adverse effects on students who are still developing critical thinking skills. This underscores the importance of balancing the use of ChatGPT with the need to maintain and cultivate independent thought and learning processes.
\begin{quote}
    \textit{``I only used it [ChatGPT] to do some coding or refine my writing, which probably doesn't matter. However, say if a secondary or high school student use it for coursework, then they probably won't learn anything due to the lack of their thinking process.''} (P9)
\end{quote}

\subsection{Incidental Learning through ChatGPT}
It has been well-established that learning can happen in neither structured nor classroom-based environments such as workplace \cite{marsick2015informal}. Through the use of ChatGPT to perform daily tasks and tackle problems, our participants have demonstrated the informal and incidental learning process happened via ChatGPT.

\subsubsection{ChatGPT Handling Email Communications}
Participants mentioned that they extensively use ChatGPT for tasks that are \textit{``lacking in creativity or high in repetitiveness,''} such as drafting emails, weekly reports, speeches, and the like. For example, P14, a teaching assistant (TA), uses ChatGPT to accelerate her process of responding to student emails due to the large number of student inquiries. 
\begin{quote}
    \textit{``A TA has to answer numerous student questions and respond to many emails. In fact, I use ChatGPT extensively for replying to emails. After using it frequently, you start to notice the vocabulary used, and you can learn a bit from it. Eventually, you might not even need ChatGPT anymore. You'll be able to write on your own, which I think is very good.''} (P14) 
\end{quote}

While P14 might not intend to learn knowledge through writing e-mails, she mentioned that with the interaction with ChatGPT, she has learned some writing techniques from the process, as she started to \textit{``notice the vocabulary''}.

\subsubsection{ChatGPT as a Professional Writing Assistant}
Another major use of ChatGPT frequently mentioned by participants is as an assistant to refine their writing for professional purposes. For instance, P10 often uses ChatGPT as a translator to look up information in other languages, stating, \textit{``ChatGPT can quickly translate content into another language, and the quality of translated content is quite good.''} P3, on the other hand, uses ChatGPT as a tool for enhancing grammar in conjunction with other software. She explained, \textit{``After using Grammarly, I sometimes feel that the sentences aren’t very authentic. Then, I put the whole paragraph in ChatGPT to polish it, and I find that the changes ChatGPT makes are particularly good, very authentic.''} In this case, P3 explicitly compared ChatGPT's performance with other professional tools in educational contexts such as Grammarly, and felt ChatGPT was more authentic. Simiarly to the email communication scenario, the use of ChatGPT for translating and refining text can incidentally enhance users' language proficiency. However, this incidental learning process may not naturally happen in every usage scenario. P1, an undergraduate student, expresses concerns about dependency and laziness stemming from using ChatGPT. Instead, he strongly supports the use of ChatGPT in tasks that are highly repetitive because they are not meant for enhancing one's knowledge or skills.

\subsection{ChatGPT Sparked Ethical Consideration in Coursework for both Students and Teachers} 
Participants generally viewed LLMs favorably, acknowledging its strengths in rapid data processing, efficiency in providing overviews or summaries, generating preliminary insights, and user-friendly format. During the interviews, we learned that the participants had already been using or encountering tasks handled by LLMs in their work or studies to varying extents. For instance, P14 noticed articles generated directly by ChatGPT in classroom assignments and expressed concerns about intellectual property and ethics, stating, \textit{``You [the student] need to rely on yourself to complete the content for the assignment to be truly effective. It [the assignment] should be a product of labor.''} As an educational professional, she implemented measures to restrict the use of ChatGPT, \textit{``This year, after discussing with my lecturer, we [decided to] directly prohibit students from using ChatGPT in the entire class.''}

However, although most participants expressed a certain level of concern, even implementing restrictions on the use of ChatGPT in educational settings, as P4 noted: \textit{``This trend is inevitable... there's no way to stop students from using these technologies [LLMs-based application].''} It's clear that the concerns are not against LLM tools per se but rather about how they are used. Even though there are potential ethical risks in using ChatGPT in educational settings, at the end of the interviews, P14 also expressed interest in and suggested a course at the school on how to harness ChatGPT, discussing \textit{``how to use prompts, as well as some limitations or advantages of the application itself, or even the current progress, these are all things that can be included in a course.''} In addition, all participants in the interview study expressed interest in understanding how to use ChatGPT better and how to design prompts more effectively. They also hope schools or businesses can offer specialized courses related to ChatGPT or integrate them into the curriculum as a part of assignments and teaching.

\section{Discussion}
In our investigation into the integration of ChatGPT in educational settings, we first delved deeply into how participants are utilizing this tool and what their concerns are. We find that ChatGPT is predominantly used for a variety of purposes, such as general information inquiries, literature reviews, content creation, language refining, data organization, and inspiration in content summarization. We delve into the dilemmas faced by students in their interactions with ChatGPT, exploring the dual aspects of efficiency and trust. This includes the complexities of their engagement with these advanced artificial intelligence tools, highlighting the inherent potential and pitfalls in the use of ChatGPT. The discussion not only reflects the evolving relationship between humans and AI in the context of education but also contemplates the broader implications of this interaction for knowledge acquisition, critical thinking, and ethical considerations in the era of rapidly advancing AI technologies. 

\subsection{Student's Dilemmas in Collaborating with LLMs: Efficiency and Trust Crisis}

From the interview results, participants show great interest in using LLMs-based applications but experience a sense of distrust during their use. This is primarily due to the ``black box'' issue inherent in LLMs~\citep{8466590}. Applying ``explanatory AI'' is considered a right choice~\citep{10.1145/3538392}. In response to this, on one hand, researchers are continually enhancing the explainability of LLMs through methods like modeling design~\citep{joshi2023repair}, ``jailbreaking''~\citep{chao2023jailbreaking} or prompt engineering~\citep{zhang2023redefining,zhang2023qualigpt}. On the other hand, we observed that participants' behavior when utilizing LLM applications tends to be more utilitarian, paralleling the widely discussed challenges of data privacy and convenience in the information age~\citep{carroll2008theorizing}. In collaboration with LLMs, participants are willing to trade off the demand for the reliability of real data for seemingly usable results in exchange for higher efficiency. This trade-off can create a toxic effect; such collaborative behavior may effectively address the immediate issues but increases the transmission of incorrect information~\citep{rush2014social}, which is very dangerous in the field of education. Over time, this might exacerbate the formation of information cocoons and intensify existing biases~\citep{preiksaitis2023opportunities}.

User scrutiny might help mitigate these risks, however, reviewing and investigating the authenticity of the content generated by LLMs can lead to a significant amount of additional work~\citep{zhang2023redefining}. The personal experience of users is crucial for swiftly reviewing and discerning the authenticity of generated content, yet such rich experience is often not applicable to novices, especially for understanding tacit knowledge~\citep{10.1145/6138.6145}. If they use LLMs and receive incorrect information, novices might be unable to discern this and could be influenced by it.

Novices might need to be more cautious than experienced individuals when using LLMs. From this perspective, the role of a reviewer or mentor for LLM-generated content could emerge as a new profession or replace TAs. Educators should supervise novices using LLM-based applications, having students include their LLM use in submissions. Tutors can then evaluate these submissions, prompting students to rethink potential shortcomings in the generated results, thereby promoting critical thinking and exploring new possibilities.

Furthermore, we would like to emphasize the potential and importance of LLMs in facilitating incidental learning. Incidental learning is often described as learning that occurs unconsciously, where learning is a byproduct of another activity~\citep{marsick2001informal}. As participants have mentioned, using ChatGPT for extensive email editing can subtly instill better grammar and vocabulary. When students use LLMs to complete tasks, apart from the repetitive nature of the operation which can be seen as a form of continuous intentional learning state~\citep{hulstijn2003incidental}, this kind of writing may trigger and enhance incidental learning~\citep{watkins1992towards}. This is similar to what the participants mentioned about learning vocabulary and grammar while writing emails. Further research is needed to explore the role of LLMs in facilitating incidental learning, including analyzing how interactions with LLMs influence the subconscious acquisition of skills and knowledge. 

\subsection{Students' Concerns About More ``Intelligent'' Artificial Intelligence}

Building on the previous discussion of human trust in AI, it is essential to delve deeper into the intelligence of ChatGPT within HCI, particularly compared to traditional AI systems. ChatGPT's intelligence is multifaceted, encompassing various components that highlight its uniqueness in HCI. Beyond performance, it is crucial to examine its explainability and interactive capabilities.

Firstly, ChatGPT's intelligence is evident in its ability to understand and generate natural language, surpassing the simple keyword matching of traditional search engines. ChatGPT can contextually analyze queries, understand nuances, and provide targeted responses, thanks to extensive training on vast text data. However, this raises concerns about potential bias and inaccuracies, unlike traditional AI systems that rely on structured and vetted data~\citep{li-etal-2023-halueval}. Issues of fabricated facts in LLMs can be partially mitigated by prompt engineering but still require human review~\citep{zhang2023redefining}. Users generally trust traditional search engines like Google over LLMs because ChatGPT's advanced language processing creates a semblance of understanding and empathy~\citep{bang2023multitask}, often leading to the attribution of ``human-like'' qualities to the AI~\citep{augenstein2023factuality}. Because, humans have the ability to interpret external realities but cannot follow instructions~\cite{sartori2022sociotechnical}. The ``human-like'' communication of LLMs' applications weakens the concept of these AIs following instructions.

With a ``mindful brain'' that traditional algorithm-based software lacks~\cite{edelman1982mindful}, search engines are able to offer results that garner greater trust from humans. They achieve this by structuring and filtering data in a way that aligns with human social experiences, primarily through mechanisms like keyword searching and algorithmic curation. This process delivers facts in an emotionless, factual manner. Even though the ranking of results provided by search engines may be influenced by recommendation algorithms, the results themselves are not considered to be fabricated by AI.

For experienced users, ChatGPT’s outputs are used cautiously and selectively. However, our preliminary results indicate that participants are concerned about the potential for ChatGPT’s interaction style to foster lazy thinking. Students, in particular, may lack the ability to accurately judge the content, leading to dependence and impaired judgment, similar to the effects of alcohol.

To effectively utilize LLM-based applications, they must produce transparent and traceable results, necessitating heightened vigilance from users. ChatGPT represents a significant AI advancement, requiring thorough examination from ethical, epistemological, and temporal perspectives to ensure integration into human knowledge frameworks. Despite its capabilities, there remains a hesitance to fully trust advanced AI systems. Ensuring the responsible and transparent use of such technologies is essential to maintain information integrity in the AI era.


From a learner's perspective, ``human-like'' AI systems might be seductive yet potentially harmful. Trust in ChatGPT due to its interactive capabilities could lead to the spread of incorrect knowledge. As noted by Preiksaitis and Rose~\citep{preiksaitis2023opportunities}, in rapidly evolving educational fields, knowledge is always outdated. Hence, AI can trigger deeper thinking, but students must approach AI critically and skeptically to navigate its imperfections in accuracy and reliability.

\subsection{Limitations and Future Work}
There are some limitations to our study. Firstly, our participants were mainly PhD students because they are more familiar with technology and ChatGPT in general,  with advanced knowledge in education and advanced skills in novel technology, so they may not be representative of undergraduates in education.  Technology literacy is important factors that can shape students' behaviors and perception, undergraduates with low literacy or just start to use chatgpt may have different perceptions. Future work should extend our finding to diverse student groups based on education backgrounds such as levels, majors, programs.  Secondly, the study mentioned some interesting points that were not covered in depth, such as copyright and ambiguity differentiation between AI and human-generated content, which could potentially lead to misinformation spreading and knowledge erosion. Future work should explore the potential disruption caused by LLM in the education context and try to mitigate these negative impacts in advance with management and technological infrastructure development and education pedagogy design.

\section{Summary}
We have summarized some valuable future work from the results and discussion sections, such as (1) the impact of ``human-like'' AI on students' judgment, as mentioned in the discussion section. However, the ubiquity and specific severity of such impacts in the educational field remain unclear. Future studies could investigate aspects like the degree of students' dependence on LLM-based applications in education and how students' judgment and learning methods are affected by LLM-based applications. (2) We suggest increasing human tutor or supervisor involvement to monitor and guide students’ use of ChatGPT. Future research may provide more specific training programs, as well as address more specific questions like who is more suitable to act as a ChatGPT supervisor in an educational environment.  Is it teachers with expertise in a particular academic field? Teachers with a background in ChatGPT research? student TAs? Or another AI?

\begin{acks}
This work was supported by the Center for Socially Responsible Artificial Intelligence (CSRAI) at Pennsylvania State University. This work is part of the ``Optimizing Large-Scale Language Model-Based AI Integration and Human-Computer Interaction in Educational Scenarios'' project, funded by the Big Ideas Grant (BIG) Summer 2023, CSRAI. We would like to extend our thanks to all participants for their valuable involvement.
\end{acks}

\bibliographystyle{ACM-Reference-Format}
\bibliography{reference}



\end{document}